\documentclass[letterpaper,twocolumn,english,aps,prl,floatfix,showpacs,tightenlines]{revtex4}
\usepackage[T1]{fontenc}
\usepackage[cp1250]{inputenc}
\usepackage{amsmath}
\usepackage{amssymb}
\usepackage{graphicx}
\usepackage{esint}

\makeatletter

\pdfpageheight\paperheight
\pdfpagewidth\paperwidth

\@ifundefined{textcolor}{}
{%
 \definecolor{BLACK}{gray}{0}
 \definecolor{WHITE}{gray}{1}
 \definecolor{RED}{rgb}{1,0,0}
 \definecolor{GREEN}{rgb}{0,1,0}
 \definecolor{BLUE}{rgb}{0,0,1}
 \definecolor{CYAN}{cmyk}{1,0,0,0}
 \definecolor{MAGENTA}{cmyk}{0,1,0,0}
 \definecolor{YELLOW}{cmyk}{0,0,1,0}
 }

\newcommand{\hc}{\mathrm{h.c.}}
\newcommand{\tr}{\mathrm{tr}}

\newcommand{\1}{\leavevmode{\rm 1\ifmmode\mkern  -4.8mu\else\kern -.3em\fi I}}

\usepackage{babel}

\usepackage{hyperref}
\usepackage{times}
\usepackage{mciteplus}

\makeatother

\usepackage{babel}
\begin{document}

\title{Gaussian Equilibration}

\author{Lorenzo Campos Venuti and Paolo Zanardi}

\affiliation{Department of Physics and Astronomy and Center for Quantum Information
Science \& Technology, University of Southern California, Los Angeles,
California 90089-0484, USA}
\begin{abstract}
A finite quantum system evolving unitarily equilibrates in a probabilistic
fashion. In the general many-body setting the time-fluctuations of
an observable $\mathcal{A}$ are typically exponentially small in
the system size. We consider here quasi-free Fermi systems where the
Hamiltonian and observables are quadratic in the Fermi operators.
We first prove a novel bound on the temporal fluctuations $\Delta\mathcal{A}^{2}$
and then map the equilibration dynamics to a generalized classical
XY model in the infinite temperature limit. Using this insight we
conjecture that, in most cases, a central limit theorem can be formulated
leading to what we call \emph{Gaussian equilibration}: observables
display a Gaussian distribution with relative error $\Delta\mathcal{A}/\overline{\mathcal{A}}=O(L^{-1/2})$
where $L$ is the dimension of the single particle space. The conjecture,
corroborated by numerical evidence, is proven analytically under mild
assumptions for the magnetization in the quantum XY model and for
a class of observables in a tight-binding model. We also show that
the variance is discontinuous at the transition between a quasi-free
model and a non-integrable one. 
\end{abstract}


\maketitle

\section{Introduction}

Out of equilibrium and equilibration dynamics of closed quantum systems
have been recently at the center of a renewed and intense interest
\citep{kinoshita_quantum_2006,calabrese_time_2006,hofferberth_non-equilibrium_2007,rigol_thermalization_2008,linden_quantum_2009,campos_venuti_unitary_2010,campos_venuti_exact_2011}.
In particular the issue whether quantum integrability plays a key
role in equilibration and, if so, which one, has been investigated
by several authors \citep{rigol_relaxation_2007,kollath_quench_2007,barthel_dephasing_2008}.
In this paper we will address this problem along the lines of the
probabilistic approach to quantum equilibration advocated in \citep{campos_venuti_unitary_2010,campos_venuti_universality_2010,campos_venuti_exact_2011}.
Here the central object is the (infinite time) full time statistics
of the expectation value of a quantum observable. We will focus on
a particular, yet very important class of quantum integrable systems:
quasi-free fermionic systems, i.e.~systems where both the Hamiltonian
and the observable are quadratic in the canonical Fermi operators.
Using fairly general central-limit type arguments as well as explicit
analytic examples we will argue that a sharp distinctive feature of
these systems, as opposed to the general interacting ones, is an exponential
enhancement of the amplitude of the temporal fluctuations of a quadratic
observable around its mean value. This appears to be a precise and
quantitative way to make sense of the common folklore that integrability
leads to a poorer (or no) equilibration.

The system is initialized in a generic state $\rho_{0}$ with $N$
particles and both the evolution Hamiltonian $H$ and the observable
$A$ are quadratic in the fermionic operators %
\footnote{Throughout the paper we use interchangeably the terms quasi-free or
quadratic, for observables quadratic in Fermi operators. %
}. The Hamiltonian is $H=\sum_{x,y}c_{x}^{\dagger}M_{x,y}c_{y}=c^{\dagger}Mc$
(notation $c^{\dagger}=\left(c_{1}^{\dagger},\ldots,c_{L}^{\dagger}\right)$,
$L$ number of sites). 
 The general quadratic observable has the form $A=\sum_{x,y}c_{x}^{\dagger}a_{x,y}c_{y}=c^{\dagger}ac$.
We will assume that $\left\Vert a\right\Vert _{\infty}=O\left(1\right)$
\footnote{To achieve this, for example, in the translation invariant case, it
suffices to have $a_{x,y}=a\left(x-y\right)$ sufficiently fast decaying. %
} as this guarantees that the expectation values of $A$ scales at
most extensively with the system size %
\footnote{By diagonalizing $a$ and exploiting unitary invariance of the operator
norm, one finds $\langle A\rangle\le\|A\|=\|\sum_{\mu}\alpha_{\mu}c_{\mu}^{\dagger}c_{\mu}\|\le\sum_{\mu}|\alpha_{\mu}|\|c_{\mu}^{\dagger}c_{\mu}\|=\sum_{\mu}|\alpha_{\mu}|\le\|a\|_{\infty}L$.
Here the $\alpha_{\mu}$'s are the eigenvalues of $A$ and the $c_{\mu}$'s
the fermionic operators associated to the corresponding eigenvectors. %
}. 
The main object of investigation is $\mathcal{A}\left(t\right)=\tr\left(Ae^{-itH}\rho_{0}e^{itH}\right)$.
Exploiting the quadratic nature of the problem and introducing the
covariance matrix $R_{y,x}:=\tr\left(\rho_{0}c_{x}^{\dagger}c_{y}\right)$
($0\le R\le\1$) one can show that the expectation value $\mathcal{A}\left(t\right)$
reduces to a trace in the one-particle space: 
\begin{equation}
\mathcal{A}\left(t\right)=\tr\left(ae^{-itM}Re^{itM}\right)\,.\label{eq:A_free}
\end{equation}
 Eq.~(\ref{eq:A_free}) is perfectly analogous to its many-body version
$\mathcal{A}\left(t\right)=\tr\left(Ae^{-itH}\rho_{0}e^{itH}\right)$
with $R$ playing the role of the initial state $\rho_{0}$. There
is however one importance difference: while $\tr\rho_{0}=1$ one has
$\tr R=N=\nu L$ , i.e.~is \emph{extensive} (we defined $\nu=N/L$
the filling factor).

For unitary evolution in finite systems, the density matrix $\rho\left(t\right)=e^{-itH}\rho_{0}e^{itH}$
does not converge neither in the strong nor in the weak topology \citep{campos_venuti_unitary_2010}.
Equilibration must be formulated in probabilistic terms. Given the
observation time window $\left[0,T\right]$, the observable $\mathcal{A}\left(t\right)$
has probability $P_{A}\left(\alpha\right)d\alpha$ of being in the
interval $\left[\alpha,\alpha+d\alpha\right]$, where the probability
density is given by $P_{A}\left(\alpha\right)=\overline{\delta\left(\mathcal{A}\left(t\right)-\alpha\right)^{T}}$,
and the time average operation is $\overline{f^{T}}=T^{-1}\int_{0}^{T}f\left(t\right)dt$.
For simplicity we will always take the limit $T\to\infty$ when taking
time average and write simply $\overline{f}$ in place of $\overline{f^{\infty}}$.
Roughly speaking an observable \emph{$A$ equilibrates} if its probability
density $P_{A}\left(\alpha\right)$ is highly peaked around its mean
$\overline{\mathcal{A}\left(t\right)}=\tr\left(A\overline{\rho}\right)$
\citep{campos_venuti_unitary_2010}. The role of equilibrium state
is played by the time-averaged density matrix $\overline{\rho}.$

The question we are going to address here is: what is the size of
the fluctuations of $P_{A}\left(\alpha\right)$ for observables $A$
in this quasi-free setting?

Before tackling this question and concentrate on fluctuations, let
us point out a few remarks on the nature of the equilibrium state
itself. 
If the spectrum is non-degenerate, the average, dephased, state $\overline{\rho}$
has the form $\overline{\rho}=\sum_{n}p_{n}|n\rangle\langle n|$ where
$|n\rangle$ are many-body eigenstate of $H$ corresponding to energy
$E_{n}$ and $p_{n}=\langle n|\rho_{0}|n\rangle$. The powers of the
Hamiltonian $H^{n}$, $n=0,1,\ldots,d-1$ are linearly independent
if there are $d$ different eigenvalues $E_{n}$ and if the spectrum
is non-degenerate $d$ coincides with the Hilbert space dimension
\footnote{Note that one can have a non-degenerate spectrum even in free systems.
It suffices, e.g., to have a rational independent one-particle spectrum.%
}. In this latter case the average state can always be written in the
form $\overline{\rho}=\exp\left[\sum_{k=0}^{d-1}\alpha_{k}H^{k}\right]$,
the so called GGE (generalized Gibbs ensemble) \citep{rigol_relaxation_2007}.
The coefficients $\alpha_{n}$ depend on the initial state $\rho_{0}$
and on the eigenvectors $\{|n\rangle\}$. The condition to write the
coefficients $\alpha_{n}$ in terms of the $p_{n}$ is precisely that
of the invertibility of the Vandermonde matrix $V_{n,k}=\left(E_{n}\right)^{k-1}$
($n,k=1,\ldots,d$), i.e., once again, non-degeneracy of the spectrum,
since $\det V=\prod_{n<k}\left(E_{k}-E_{n}\right)$. The relation
expressing the $p_{n}$'s in terms of the $\alpha_{k}$'s is: $p_{n}=\exp\left[\sum_{k=0}^{d-1}\alpha_{k}E_{n}^{k}\right]$.
Since $\overline{\rho}=\exp\sum_{n}\ln p_{n}|n\rangle\langle n|=\exp\sum_{k=0}^{d-1}\sum_{n}\alpha_{k}E_{n}^{k}|n\rangle\langle n|$
, the inverse relation is, in vector notation $\boldsymbol{\alpha}=V^{-1}\ln\boldsymbol{p}$.
The inverse of the Vandermonde matrix can be found explicitly by expanding
the identity $|n\rangle\langle n|=\prod_{k\neq n}\frac{H-E_{k}}{E_{n}-E_{k}}$,
multiplying by $\ln p_{n}$, summing over $n$ and exponentiating
see %
\footnote{Explicitly $\sum_{n}p_{n}|n\rangle\langle n|=\exp\left[\sum_{n}\sum_{k=0}^{d-1}\frac{\gamma_{k,n}}{\Delta_{n}}\ln p_{n}\, H^{k}\right]$
where $\gamma_{k,n}=\left(-1\right)^{d-1-k}\sum_{j_{1}<\cdots<j_{d-k};j_{i}\neq n}E_{j_{1}}\cdots E_{j_{d-k}}$~,
and $\Delta_{n}=\prod_{j\neq n}\left(E_{n}-E_{j}\right)$.%
}

Let us now go back to the context of Eq.~(\ref{eq:A_free}). Since
$0\le R\left(t\right)\le\1$ the time averaged covariance $\overline{R}=\overline{e^{-itM}Re^{itM}}$
satisfies $0\le\overline{R}\le\1$ and so defines a Gaussian state
$\rho_{\overline{R}}$ with covariance $\overline{R}$. Moreover,
since $\tr R\left(t\right)=N$, for all $t$, $\tr\overline{R}=N$,
$\rho_{\overline{R}}$ is a Gaussian state with $N$ particles. Now,
for what concerns quadratic observables of the kind $A=c^{\dagger}ac$,
their time average expectation value is the same as that obtained
with $\rho_{\overline{R}}$: $\overline{\mathcal{A}}=\tr a\overline{R}=\tr A\rho_{\overline{R}}$.
In other words the states $\overline{\rho}$ and $\rho_{\overline{R}}$
\emph{are the same} when restricted to quadratic observables. A generic
Gaussian state can be written in the form $\rho_{R}={\cal N}\exp\left(c^{\dagger}G(R)c\right)$
where $G(R):=\log[R(1-R)^{-1}]$ \citep{peschel_calculation_2003}
and ${\cal N}$ is a normalization constant. From this it immediately
follows that $\rho_{\overline{R}}$ can be written as $\rho_{\overline{R}}={\cal N}\exp\sum_{k}\lambda_{k}c_{k}^{\dagger}c_{k}$
where $c_{k}$'s are eigenmodes of $H$ and coincides with equation
(8) of \citep{rigol_relaxation_2007}. When the system equilibrates
i.e., ${\cal A}(t)\to\tr\left(A\rho_{\overline{R}}\right)$ this remark
shows the validity of the GGE for \emph{any} initial state $\rho_{0}$
and \emph{all quadratic} observables. Indications that $\overline{\rho}$
converges to $\rho_{\overline{R}}$ in some sense as the size increases
were indeed already present in the literature. For instance in \citep{lanford_approach_1972}
it was shown that, in the thermodynamic limit, $\rho\left(t\right)\to\rho_{\overline{R}}$
weakly as $t\to\infty$, while in \citep{cramer_exact_2008,cramer_quantum_2010}
a particular form of strong convergence was derived, when considering
subsystems (though for a Bosonic system).

\section{A bound on the variance}

Assuming the non-resonant condition on the energies ($E_{n}-E_{m}=E_{p}-E_{q}$
implies $n=m$ and $p=q$ or $n=p$ and $m=q$), Reimann has shown
\citep{reimann_foundation_2008} that the temporal fluctuations $\Delta\mathcal{A}^{2}=\overline{\left(\mathcal{A}\left(t\right)-\overline{\mathcal{A}}\right)^{2}}$
satisfy $\Delta\mathcal{A}^{2}\le\mathrm{diam}\left(A\right)^{2}\tr\overline{\rho}^{2}$
where $\mathrm{diam}\left(A\right)$ is the maximum minus the minimum
eigenvalue of $A$. Now in general, for most initial states $\rho_{0}$,
the purity $\tr\overline{\rho}^{2}$ is exponentially small in the
system size implying exponentially small fluctuations. Here is one
argument. First note that $\tr\overline{\rho}^{2}=\overline{\mathcal{L}\left(t\right)}$
with $\mathcal{L}\left(t\right)$ being the Loschmidt echo: $\mathcal{L}\left(t\right)=\left|\langle\psi_{0}|e^{-itH}|\psi_{0}\rangle\right|^{2}$
which admits the following cumulant expansion $\mathcal{L}\left(t\right)=\exp\left[2\sum_{n=1}^{\infty}\left(-t^{2}\right)^{n}\langle H^{2n}\rangle_{c}/\left(2n\right)!\right]$
\citep{campos_venuti_unitary_2010}. Here $\langle H^{2n}\rangle_{c}$
are the cumulants of $H$ computed with $\rho_{0}$. The point is
that if $\rho_{0}$ is sufficiently clustering, but not an eigenstate
of $H$, all the cumulants are extensive in the system size and non-zero.
Then $\mathcal{L}\left(t\right)=\exp\left[-L^{D}g\left(t\right)+\mathrm{corrections}\right]$
where $g\left(t\right)\ge0$ does not depend on the size and, for
sufficiently large $L$ one has $\overline{\mathcal{L}}\le\exp\left[-L^{D}\min_{t}g\left(t\right)\right]$
\footnote{When $g\left(t\to\infty\right)$ exists, for large $L$ one can use
the more precise estimate $\overline{\mathcal{L}}\simeq\exp\left[-L^{D}g\left(\infty\right)\right]$%
}. 

In the quasi-free setting the non-resonant condition does not hold.
Let us then seek for the analogous of the bound of Reimann in our
quasi-free case. Let the one-particle Hamiltonian have the following
diagonal form $M=\sum_{k}\Lambda_{k}|k\rangle\langle k|$. The time
averaged covariance matrix is then $\overline{R}=\sum_{k}\langle k|R|k\rangle|k\rangle\langle k|$.
We also define $F_{k,q}=\langle k|a|q\rangle\langle q|R|k\rangle$.
Assuming the non-resonance condition for the one-particle spectrum,
one gets $\Delta\mathcal{A}^{2}=\tr F^{2}-\sum_{k}\left(F_{k,k}\right)^{2}\le\tr F^{2}=\sum_{k,q}\left|\langle k|a|q\rangle\right|^{2}\left|\langle q|R|k\rangle\right|^{2}$.
Now $R$ is a non-negative operator and so induces a (possibly degenerate)
scalar product which satisfies Cauchy-Schwarz inequality: $\left|\langle q|R|k\rangle\right|^{2}=\left|\langle q|k\rangle_{R}\right|^{2}\le\langle q|q\rangle_{R}\langle k|k\rangle_{R}=\langle q|R|q\rangle\langle k|R|k\rangle$.
This leads us to 
\begin{equation}
\Delta\mathcal{A}^{2}\le\tr\left(a\overline{R}a\overline{R}\right)\le\left\Vert a\right\Vert _{\infty}^{2}\tr\overline{R}^{2}\label{eq:bound_variance}
\end{equation}
 Now, since $0\le\overline{R}\le\1$, $\tr\overline{R}^{2}\le\tr\overline{R}=\tr R=N$,
we finally obtain $\Delta\mathcal{A}^{2}\le\left\Vert a\right\Vert _{\infty}^{2}\nu L\,.$
While equation (\ref{eq:bound_variance}) is the quasi-free analog
of the Reimann's bound it implies some important differences with
respect to the general (non-free) case. Consider the situation where
the observable $A$ is extensive. In the non-free case the diameter
of $A$ is extensive, i.e.~$\mathrm{diam}\left(A\right)=O\left(L^{D}\right)$
in $D$ spatial dimensions. Moreover, the minimum value of the purity
$\tr\overline{\rho}^{2}$ is $1/d$ and so is exponentially small
in the system size. 
In the quasi-free setting, instead, the minimum value of $\tr\overline{R}^{2}$
in Eq.~(\ref{eq:bound_variance}) is $\min\tr\overline{R}^{2}=N^{2}/L=\nu^{2}L$,
since $\overline{R}/N$ defines a density matrix for which the minimum
purity is the inverse of the matrix dimension $L$. All in all, recalling
that $\left\Vert a\right\Vert _{\infty}=O\left(1\right),$ for extensive
observables, the minimum of the bound to the variance is $O\left(L^{2}e^{-\gamma L}\right)$,
$\gamma>0$, (for systems of linear size $L$), whereas in the quasi-free
setting one has $O\left(L\right)$. This seems to hint at the fact
that fluctuations in the quasi-free setting are proportional to the
system size and are hence much larger than in the non-free case where
they are exponentially small in the volume. Of course Eq.~(\ref{eq:bound_variance})
is just an upper bound, and nothing prevents, in principle, from having
a much smaller variance. For example, whenever the initial state or
the observable commute with the Hamiltonian, $\mathcal{A}\left(t\right)$
is constant and its fluctuations are zero. We will always avoid such
pathological situations. In the following, instead, we will argue
that the extensive behavior of the fluctuations, in quasi-free systems,
is in fact quite general, leading to a $1/\sqrt{L}$ scaling of the
relative error $\sqrt{\Delta\mathcal{A}^{2}}/\overline{\mathcal{A}}$
for a generic observable. Indeed such a $1/\sqrt{L}$ scaling has
been observed to hold for a quadratic Hamiltonian even for more general
observables (see supplementary material of \citep{cassidy_generalized_2011})
and even in presence of disorder except for quenches into a localized
phase \citep{gramsch_dynamics_2012}.

\section{Mapping to a classical XY model}

Let us write again the generic expectation value (\ref{eq:A_free})
in the basis which diagonalizes $M$: $\mathcal{A}\left(t\right)=\overline{\mathcal{A}}+2\sum_{k<q}\left|F_{k,q}\right|\cos\left(t\left(\Lambda_{k}-\Lambda_{q}\right)+\phi_{k,q}\right)$
with $\phi_{k,q}=\arg F_{k,q}$. To obtain information on the probability
density $P_{A}\left(\alpha\right)$ we consider the generating function
$\chi_{A}\left(\lambda\right):=\overline{e^{\lambda\left(\mathcal{A}\left(t\right)-\overline{\mathcal{A}}\right)}}$.
Now we observe that \emph{if the (one-particle) energies are rationally
independent (RI)}, as a consequence of a theorem on the averages,
the infinite time average of $\mathcal{A}\left(t\right)$ is the same
as the uniform average over the torus $\mathbb{T}^{L}$. In this case
the generating function $\chi_{A}\left(\lambda\right)$ is exactly
given by the partition function of the generalized, classical XY model
with energy $E\left(\boldsymbol{\theta}\right)=2\sum_{k<q}\left|F_{k,q}\right|\cos\left(\theta_{k}-\theta_{q}+\phi_{k,q}\right)$
and inverse temperature $\beta=-\lambda$. The matrix $\left|F_{k,q}\right|$
defines the lattice of the interactions while the phases $\phi_{k,q}$
give the offset from which the angles are measured. Note that the
behavior of the density $P_{A}\left(\alpha\right)$ is dictated by
$\chi_{A}\left(\lambda\right)$ in a neighborhood of $\lambda=0$
which corresponds to infinite temperature of the classical XY model.

It is not difficult to engineer a situation which exactly reproduces
the standard XY model in $D$-dimension. For example, it suffices
to consider the Hamiltonian $H=\sum_{\boldsymbol{x}}\mu_{\boldsymbol{x}}c_{\boldsymbol{x}}^{\dagger}c_{\boldsymbol{x}}$,
with $\mu_{\boldsymbol{x}}$ RI ($\boldsymbol{x}$ is a point of a
$D$-dimensional lattice), choose the observable $A=\sum_{\langle\boldsymbol{x},\boldsymbol{y}\rangle}c_{\boldsymbol{x}}^{\dagger}c_{\boldsymbol{y}}$
($\langle\boldsymbol{x},\boldsymbol{y}\rangle$ indicates nearest
neighbor) and initial state $|\psi_{0}\rangle=L^{-1/2}\sum_{\boldsymbol{x}}|\boldsymbol{x}\rangle=c_{\boldsymbol{k}=0}^{\dagger}|0\rangle$.
In this case the partition function $\mathcal{Z}$ (free-energy $\mathcal{F}$)
of the classical $D$-dimensional XY model is precisely the characteristic
function of the observable $A$: $\mathcal{Z}=\chi_{A}\left(\lambda\right)$
($\mathcal{F}=\ln\chi_{A}\left(\lambda\right)$). In fact in this
case $a_{\boldsymbol{x},\boldsymbol{y}}=\delta_{\langle\boldsymbol{x},\boldsymbol{y}\rangle}$
while $R_{\boldsymbol{x},\boldsymbol{y}}=1/L$ so that $F_{\boldsymbol{x},\boldsymbol{y}}=L^{-1}\delta_{\langle\boldsymbol{x},\boldsymbol{y}\rangle}$
which defines the nearest neighbor hyper-cubic graph.

We would like to stress here that the one-particle space has a natural
underlying geometric structure. For instance, the labels $k,q$ represent
points in momentum (real) space in a superfluid (localized) phase
and the distance $\left|k-q\right|$ is well defined. Now, when\emph{
}the matrix elements $\left|F_{k,q}\right|$ decay sufficiently fast
as $\left|k-q\right|\to\infty$ the corresponding XY model is well
defined in the thermodynamic limit, i.e.~the intensive free energy
has a limit as $L\to\infty$. This happens for instance in case $\left|F_{k,q}\right|$
decays exponentially in $\left|k-q\right|$ or if one has $\left|F_{k,q}\right|\sim1/\left|k-q\right|^{\gamma}$
with $\gamma>D$. When this is the case one has $\chi_{A}\left(\lambda\right)=\exp L^{D}f\left(\lambda\right)$
where $f\left(\lambda\right)$ is the free energy per site. Moreover,
under these conditions, one expects $f\left(\lambda\right)$ to be
analytic in the high temperature, $\lambda=0$, limit, implying that
all the cumulants of $\mathcal{A}\left(t\right)$ are extensive. From
this we immediately draw the central limit theorem (CLT): as $L\to\infty$
the variable $(\mathcal{A}\left(t\right)-\overline{\mathcal{A}})/L^{D/2}$
tends in distribution to a Gaussian with zero mean and finite variance
given by $\partial_{\lambda=0}^{2}f\left(\lambda\right)$. We call
this situation \emph{Gaussian equilibration}. It is important to stress
that one cannot have Gaussian equilibration in the non-free setting
otherwise all the cumulants would scale in the same way. Instead in
the non-free setting one expects extensive average but variance exponentially
small in the volume.

We will now further corroborate these arguments with two examples
where indeed Gaussian equilibration can be proven or shown.

\section{Quench on the quantum XY model\label{sec:Quench-on-the}}

\begin{figure}
\begin{centering}
\includegraphics[width=7cm,height=50mm]{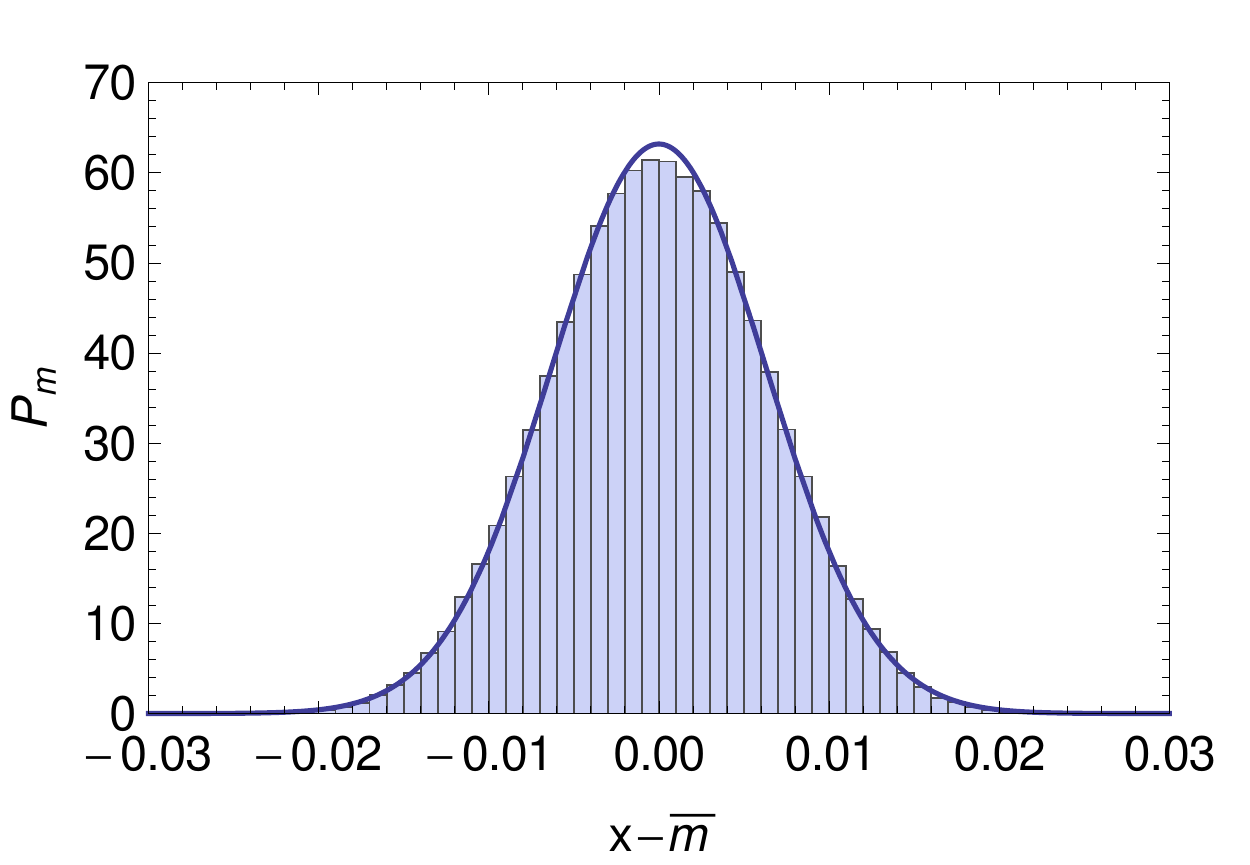} 
\par\end{centering}

\caption{Full distribution for the transverse magnetization per site $m\left(t\right)=\langle\sigma_{i}^{z}\left(t\right)\rangle$
around its mean for $L=40$. The quench is performed from ($\gamma_{0}=\gamma_{1}=1$)
$\left(h_{0}=2\right)\to\left(h_{1}=4\right)$. The histogram is obtained
sampling $m\left(t\right)$ at 200,000 random times uniformly distributed
in $[0,T_{\mathrm{max}}]$ with $T_{\mathrm{max}}=100,000$. The thick
curve is a Gaussian with zero mean and variance $\sigma^{2}=L^{-2}\sum_{k>0}W_{k}^{2}/2$
as computed in \citep{campos_venuti_unitary_2010}. \label{fig:Plot-of-m}}
\end{figure}

The Hamiltonian is given in terms of Pauli spin operators $\sigma_{i}^{x,y,x}$
(we use periodic boundary conditions) 
\begin{equation}
H=-\sum_{i=1}^{L}\left[\left(\frac{1+\gamma}{2}\right)\sigma_{i}^{x}\sigma_{i+1}^{x}+\left(\frac{1-\gamma}{2}\right)\sigma_{i}^{y}\sigma_{i+1}^{y}+h\sigma_{i}^{z}\right]\label{eq:XY_model}
\end{equation}

In the quench scenario, the initial state is the ground state $|\psi_{0}\rangle$
of the Hamiltonian with parameters $(\gamma_{0},h_{0})$. The parameters
are then suddenly changed and $|\psi_{0}\rangle$ is evolved with
the Hamiltonian corresponding to $(\gamma_{1},h_{1})$.

The model in Eq.~(\ref{eq:XY_model}) has been long used as a testbed
for the study of quantum phase transitions in many body systems, and
more recently in the realm of out-of equilibrium unitary dynamics.
See \citep{lieb_two_1961,barouch_statistical_1970} and the more recent
monumental \citep{calabrese_quantum_2012} for more details and references.
A Jordan-Wigner transformation brings Eq.~(\ref{eq:XY_model}) to
a quadratic form in Fermi operators. Since $\sigma_{i}^{z}$ in terms
of Fermi operators is $\sigma_{i}^{z}=2c_{i}^{\dagger}c_{i}-1$, the
transverse, total magnetization $M(t)=\sum_{i}\langle\sigma_{i}^{z}\left(t\right)\rangle$
is a quadratic observable. Its expectation value in the quench setting
is given by \citep{barouch_statistical_1970,campos_venuti_unitary_2010}:
$M\left(t\right)=2\sum_{k>0}\cos\vartheta_{k}^{\left(1\right)}\cos\left(\delta\vartheta_{k}\right)+\sin\vartheta_{k}^{\left(1\right)}\sin\left(\delta\vartheta_{k}\right)\cos\left(t\Lambda_{k}^{\left(1\right)}\right)$
where $\tan\vartheta_{k}^{\left(i\right)}=-\gamma_{i}\sin k/\left(h_{i}+\cos k\right)$,
$\delta\vartheta_{k}=\vartheta_{k}^{\left(1\right)}-\vartheta_{k}^{\left(0\right)}$
and $\Lambda_{k}^{\left(i\right)}=2\sqrt{\left(\gamma_{i}\sin k\right)^{2}+\left(h_{i}+\cos k\right)^{2}}$
are the one-particle energies. The quasi-momenta are quantized according
to $k=\pi\left(2n+1\right)/L$, $n=0,1,\ldots,L/2-1$. At this point
it seems quite natural to expect that the energies $\Lambda_{k}^{\left(1\right)}$
are rationally independent. Indeed one can show that the numbers $\cos(k_{n})$
$k_{n}=\pi\left(2n+1\right)/L$, $n=1,2,\ldots,(L-1)/2$ are rationally
independent for $L$ prime \citep{campos_venuti_exact_2011}. Given
the form of the dispersion we may expect that the requirement that
$L$ is prime may be removed.

Assuming rational independence of the one-particle energies $\Lambda_{k}^{\left(1\right)}$,
the corresponding classical XY model has energy $E\left(\boldsymbol{\theta}\right)=\sum_{k>0}W_{k}\cos(\theta_{k})$
with $W_{k}=\sin\vartheta_{k}^{\left(1\right)}\sin\left(\delta\vartheta_{k}\right)$.
Each classical spin '$k$' interacts with an external field along
a fixed axis with strength $W_{k}$. The partition function factorizes,
each integral over $\theta_{k}$ gives a Bessel function $I_{0}\left(\lambda W_{k}\right)$
and we obtain $\overline{e^{\lambda\left(M(t)-\overline{M}\right)}}=\exp\sum_{k>0}\ln\left(I_{0}\left(\lambda W_{k}\right)\right)$.
Clearly $M\left(t\right)-\overline{M}$ is a sum of independent random
variables, each with zero mean and variance $W_{k}^{2}/2$. 

Now, under the --quite reasonable-- assumption of rational independence
of the one-particle energies, one can prove \emph{Gaussian equilibration
}for the observable $M\left(t\right)$. More precisely one can show
that for any value of parameters $(\gamma_{0},h_{0})\neq(\gamma_{1},h_{1})$,
the variable $(M-\overline{M})/\sqrt{L}$ as $L\to\infty$ tends in
distribution to a Gaussian with zero mean and variance $\left(2\pi\right)^{-1}\int_{0}^{\pi}(W_{k}^{2}/2)\, dk$.

We can prove this result by showing that the Lyapunov condition is
satisfied so that the central limit theorem follows from Lindeberg's
theorem (see e.g.~\citep{billingsley_probability_2012}). Following
the notation of \citep{billingsley_probability_2012} we have $s_{L}^{2}=\sum_{k}\sigma_{k}^{2}=\sum_{k>0}W_{k}^{2}/2\to L\left(4\pi\right)^{-1}\int_{0}^{\pi}W_{k}^{2}\, dk$.
Then, with $\delta=1$ $\overline{\left|X_{k}\right|^{2+\delta}}=4/(3\pi)\left|W_{k}\right|^{3}$.
The Lyapunov's condition with $\delta=1$ amounts to the vanishing
of the following quantity as $L\to\infty$: 
\begin{equation}
\frac{1}{s_{L}^{3}}\sum_{k>0}\overline{\left|X_{k}\right|^{3}}\stackrel{L\to\infty}{\longrightarrow}\frac{1}{\sqrt{L}}\frac{16}{3\sqrt{\pi}}\frac{\int_{0}^{\pi}\left|W_{k}\right|^{3}dk}{\left(\int_{0}^{\pi}\left|W_{k}\right|^{2}dk\right)^{3/2}}\,.\label{eq:Lyapunov}
\end{equation}
 Indeed the RHS of Eq.~(\ref{eq:Lyapunov}) goes to zero as $L\to\infty$
since $\left|W_{k}\right|\le2$ for all $k$ and $\left|W_{k}\right|\neq0$
for almost any $k$ for $(\gamma_{0},h_{0})\neq(\gamma_{1},h_{1})$.

\paragraph*{Remark }

For small quench close to a critical point, the function $W_{k}\simeq\sin(\vartheta_{k}^{\left(1\right)})(\partial\vartheta_{k}/\partial x)\, dx$
($x$ is the quenched variable, $x=\lambda,h$), becomes highly peaked
(for instance, the peak is around $k=\pi$ close to the Ising critical
point at $h=1$ and $W_{k}/dx$ diverges as $1/k$). For finite $L$
and sufficiently small quench, few terms $W_{k}$ dominate and one
can obtain (for finite $L$) a non-Gaussian distribution. Indeed,
as discussed in detail in \citep{campos_venuti_universality_2010},
this is the case in general: for small quenches close to a quantum
critical point, observables become a sum of few independent random
variables and the distribution acquires a universal double-peaked
form.

In figure \ref{fig:Plot-of-m} we show a plot of the distribution
of the transverse magnetization per site $m\left(t\right)=M\left(t\right)/L$
for the equilibration dynamics undergoing a quantum quench. The distribution
agrees very well with a Gaussian with variance $\sigma^{2}=L^{-2}\sum_{k>0}W_{k}^{2}/2$
obtained considering $m\left(t\right)$ as a sum of independent variables
\citep{campos_venuti_unitary_2010}. This in turns shows that the
assumptions of rational independence seems to be justified or at least
that the number of relations among the frequencies is sufficiently
small as not to break the CLT. Note that $L$ is not a prime in fig.~\ref{fig:Plot-of-m}.
By only checking that the one particle spectrum is non-degenerate,
it is easy to prove that the variance is $\Delta m^{2}=L^{-2}\sum_{k>0}W_{k}^{2}/2$
\citep{campos_venuti_unitary_2010}. Since $W_{k}$ is a bounded function
this implies immediately that $\Delta m^{2}=O\left(L^{-1}\right)$,
in accordance with the Gaussian equilibration prediction.

\section{Tight binding model}

\begin{figure}
\begin{centering}
\includegraphics[width=7cm,height=50mm]{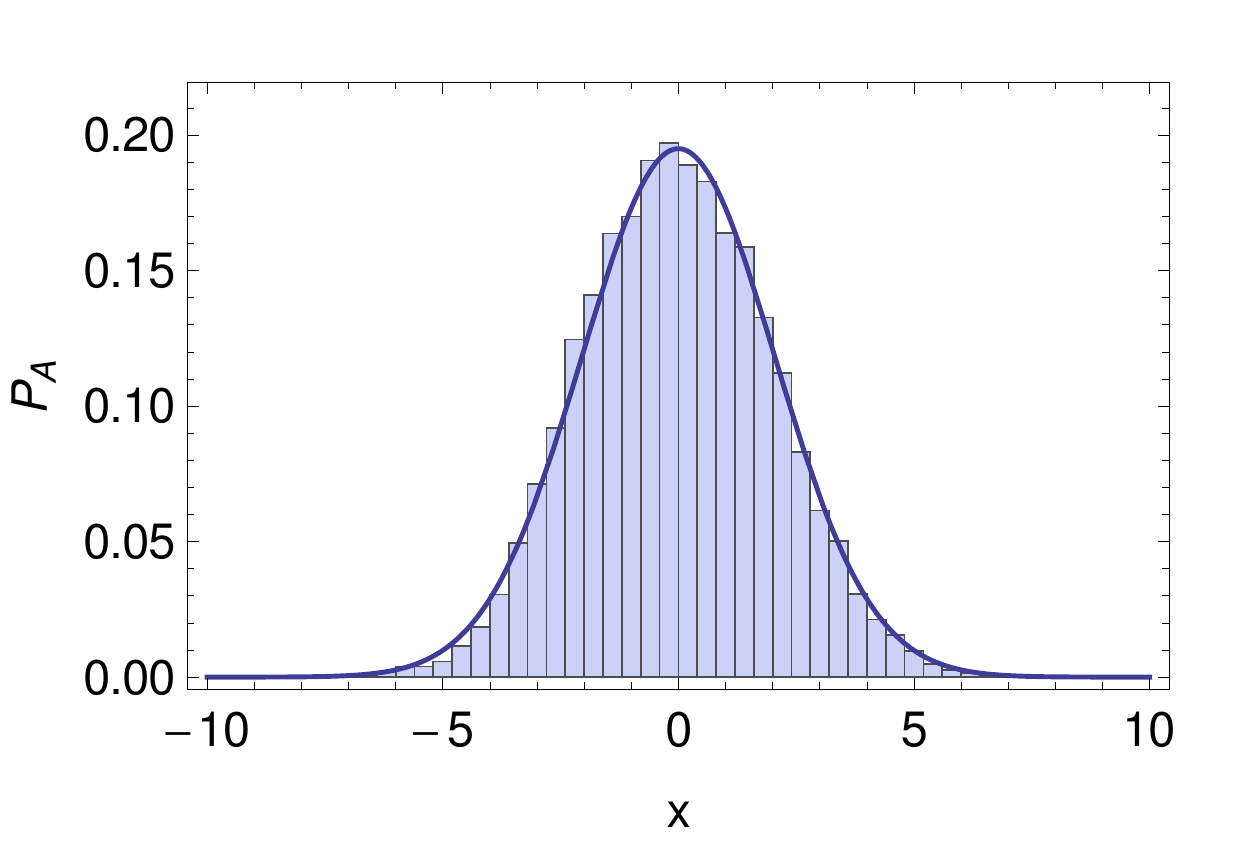} 
\par\end{centering}

\caption{Full distribution for the observable $A=\sum_{x=1}^{\ell}c_{x}^{\dagger}c_{x}$
around its mean for $L=201$, $N=100$, $\ell=101$. The histogram
is obtained sampling $\mathcal{A}\left(t\right)$ at 240,000 random
times uniformly distributed in $[0,T_{\mathrm{max}}]$ with $T_{\mathrm{max}}=180,000$.
The thick curve is a Gaussian with zero mean and variance $\sigma^{2}=0.0208L$
as computed in the text.\label{fig:Plot-of-A}}
\end{figure}

We consider here the 1-$D$ tight binding model $H=\sum_{x}\left(c_{x}^{\dagger}c_{x+1}+\hc\right)$
with twisted boundary conditions $c_{L+1}=c_{1}e^{i\theta L}$ as
proposed in \citep{tasaki_approach_2010}. As quadratic observable
we take $A$ with diagonal one-particle matrix: $A=\sum_{x=1}^{\ell}c_{x}^{\dagger}c_{x}$.
The system is initialized setting all the $N$ particles say to the
left of the chain, i.e.~the initial covariance matrix is $R=\mathrm{diag}\left(1,1,\ldots,1,0,\ldots,0\right)$
with $N$ ones and $L-N$ zeros on the diagonal. The observable $A$
is extensive for $\ell=\alpha L$, and the thermodynamic limit is
given by $\alpha=\ell/L$, $\nu=N/L$ constant and $L\to\infty$.
The time evolved observable reads $\mathcal{A}\left(t\right)=\sum_{k,q}g_{N}\left(k-q\right)g_{\ell}\left(q-k\right)e^{-it\left(\Lambda_{k}-\Lambda_{q}\right)}$
where the function $g_{\ell}\left(\xi\right)$ is given by $g_{\ell}\left(\xi\right)=L^{-1}\sum_{x=1}^{\ell}e^{-ix\xi}$.
The matrix $F_{k,q}$ depends only on the difference $k-q$: $F_{k,q}=f\left(k-q\right)=g_{N}\left(k-q\right)g_{\ell}\left(q-k\right)$.
The eigen-energies are given by $\Lambda_{k}=2\cos\left(k+\theta\right)$
and the quasimomenta can be considered quantized as $k=2\pi n/L$,
$n=0,1,\ldots,L-1$. For $\theta=0$ the energies are degenerate as
$\Lambda_{k}=\Lambda_{\pi-k}$ but for most of the $\theta$, $\pi-k$
does not belong to the Brillouin zone and the energies are non-degenerate.
In this case the average is $\overline{\mathcal{A}}=\sum_{k}g_{N}\left(0\right)g_{\ell}\left(0\right)=L\nu\alpha$.
Still the $\Lambda_{k}$ cannot be RI as $0=\tr M=\sum_{k}\Lambda_{k}$,
however it is likely that there are few relations among the energies.
In fact Tasaki showed \citep{tasaki_approach_2010} that for $L$
odd and for most of the $\theta\in(0,\pi/L)$, the one-particle energies
satisfy the non-resonant condition. This implies that, for most $\theta$,
the variance is given by 
\[
\Delta\mathcal{A}^{2}=\sum_{k,q}\left|g_{N}\left(k-q\right)g_{\ell}\left(q-k\right)\right|^{2}-\sum_{k}\left[g_{N}\left(0\right)g_{\ell}\left(0\right)\right]^{2}.
\]

The double sum above can be evaluated going back to real space with
result $L^{-2}\left[N\ell^{2}-\ell\left(\ell^{2}-1\right)/3\right]$
(assuming $N\ge\ell$ otherwise swap $N$ with $\ell$). This proves
analytically that, besides the average, also the second cumulant --the
variance-- is extensive and in particular one has $\Delta\mathcal{A}^{2}=L\left(\nu\alpha^{2}-\frac{\alpha^{3}}{3}-\nu^{2}\alpha^{2}\right)+O\left(L^{-1}\right)$.

On the other hand, using methods borrowed from statistical mechanics,
we can get an approximate analytic expression for the whole cumulant
generating function. In this approximation all the cumulants will
turn out to be extensive implying Gaussian equilibration for $\mathcal{A}\left(t\right)$
(see figure \ref{fig:Plot-of-A}). We first assume rational independence
of the energies $\Lambda_{k}$ so that the problem is mapped to an
classical XY model. A single (or a few) relation among the energies
is a sort of boundary condition for the classical model and is not
expected to change the leading, bulk, term. The energy of the classical
model for the translation invariant case is $E\left(\boldsymbol{\theta}\right)=2\sum_{d=1}^{(L-1)/2}\left|f\left(d\right)\right|\sum_{x=1}^{L}\cos\left(\left(\tilde{\theta}_{x+d}-\tilde{\theta}_{x}\right)+\phi_{d}\right)$
($\phi_{d}=\arg f\left(d\right)$ and $\tilde{\theta}_{x}$ is a periodic
extension of $\theta_{x}$, i.e.~$\tilde{\theta}_{x+nL}=\tilde{\theta}_{x}$).
Now we note that $f\left(d\right)$ is highly peaked around $d=0$,
so we approximate the energy keeping only the nearest neighbor term,
i.e.~$E\left(\boldsymbol{\theta}\right)\approx2\left|f\left(1\right)\right|\sum_{x=1}^{L}\cos\left(\left(\tilde{\theta}_{x+1}-\tilde{\theta}_{x}\right)+\phi_{1}\right)$.
This is precisely a one-dimensional (classical) XY model with periodic
boundary condition (and off-set $\phi_{1}$) and can be solved via
transfer matrix method \citep{mattis_transfer_1984}. The partition
function becomes $\mathcal{Z}=\tr\mathcal{K}^{L}$ where the transfer
matrix operator is $[\mathcal{K}h(\theta_{x})](\theta_{x+1})=\int_{0}^{2\pi}d\theta_{x}e^{\lambda\left|f\left(1\right)\right|\cos\left(\tilde{\theta}_{x+1}-\tilde{\theta}_{x}+\phi_{1}\right)}h(\theta_{x})/(2\pi)$.
The transfer matrix is non-hermitian because of $\phi_{1}$. Using
the identity $e^{K\cos\left(\alpha-\beta\right)}=\sum_{p\in\mathbb{Z}}I_{p}\left(K\right)e^{ip\left(\alpha-\beta\right)}$
where $I_{p}$ are Bessel functions (satisfying $I_{0}>I_{1}>I_{2}\cdots$
and $I_{p}\left(K\right)=I_{-p}\left(K\right)$) one sees that plane
waves $e^{im\vartheta_{x}}$ are eigenfunction of $\mathcal{K}$ with
eigenvalue $I_{m}\left(2\left|f\left(1\right)\right|\lambda\right)e^{-im\phi}$.
The largest eigenvalue in modulus, with $m=0$, gives immediately
the free energy in the large size limit: $\mathcal{F}=\ln\mathcal{Z}=L\ln\left(I_{0}(\lambda\left|f(1)\right|\right)$.
As expected, in this approximation, the cumulant generating function
is extensive and analytic in $\lambda=0$. One then has again the
CLT in its standard form: the variable $(\mathcal{A}\left(t\right)-\overline{\mathcal{A}})/\sqrt{L}$,
tends in distribution to a Gaussian in the thermodynamic limit with
variance $\partial_{\lambda=0}^{2}\mathcal{F}/L=\left|f(1)\right|^{2}/2$.
The Gaussian prediction is clearly confirmed by a numerical experiment
see fig.~\ref{fig:Plot-of-A}.

\section{Fluctuations and integrability}

\begin{figure}
\noindent \begin{centering}
\includegraphics[width=78mm,height=64mm]{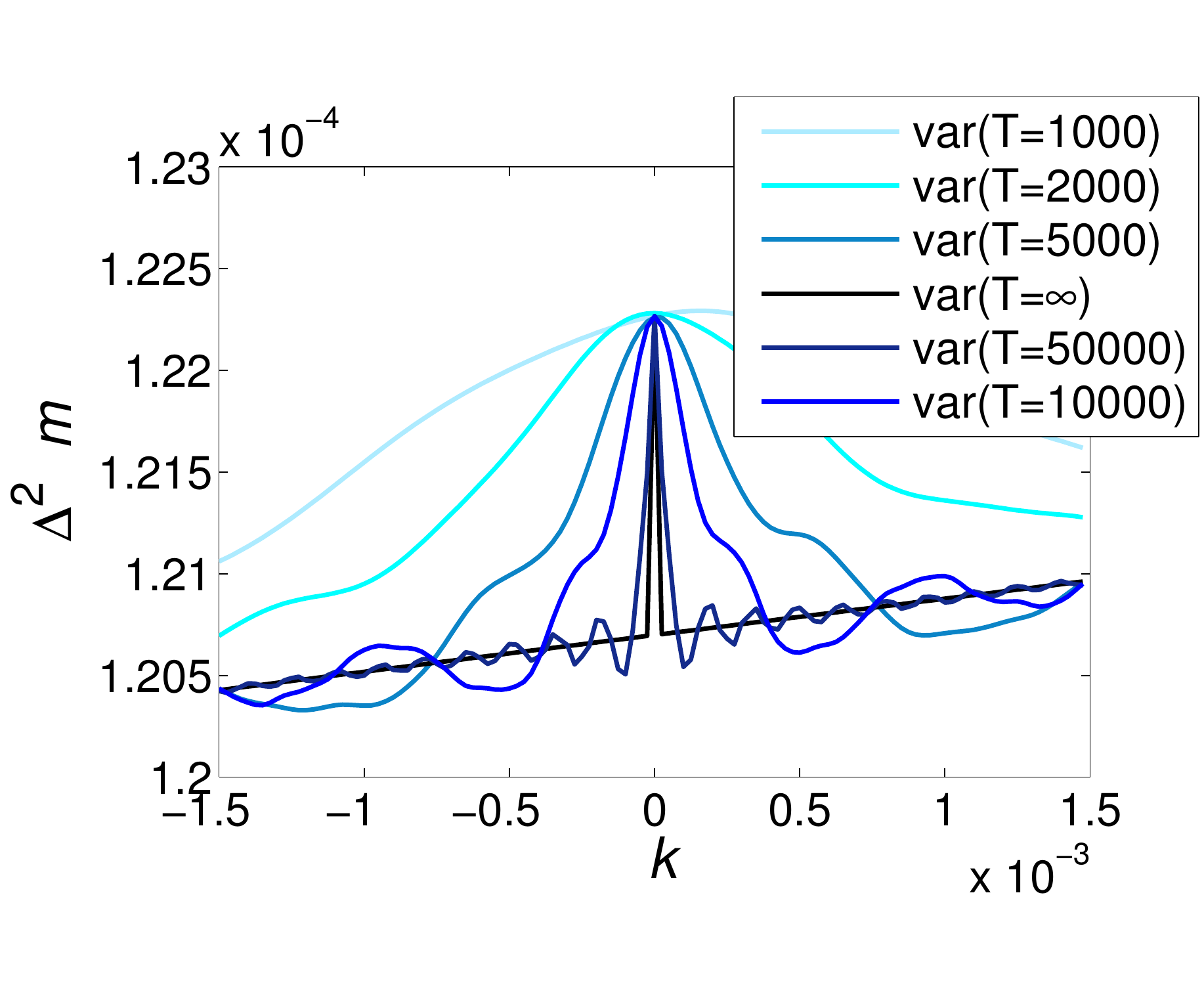}
\par\end{centering}

\caption{Variance of  $m\left(t\right)=\langle\sigma_{i}^{z}\left(t\right)\rangle$
of as a function of the integrability breaking parameter $\kappa$.
The quench parameters are ($\gamma=1$) $(\kappa_{0}=0,h_{0}=2.0),\,\to(\kappa_{1}=\kappa,\, h_{1}=2.7)$.
The size is $L=8$. Data are obtained via full diagonalization of
the Hamiltonian. \label{fig:var_kappa} }
\end{figure}

From the previous discussion it appears that the variance $\Delta\mathcal{A}^{2}$,
for a quadratic observable $\mathcal{A}$, can be used as an effective
tool for the characterization of integrable-non-integrable transition,
at least in the case where integrable models are identified with quadratic
systems. In particular we expect an increase of $\Delta\mathcal{A}^{2}$
when the evolution Hamiltonian crosses a quasi-free point. To check
this conjecture let us add a next-nearest neighbor interaction term
to Eq.~(\ref{eq:XY_model}) (for simplicity we set $\gamma=1$) which
is known to break integrability (see e.g.~\citep{beccaria_density-matrix_2006}
and references therein). Hence we consider the model 
\[
H=-\sum_{i=1}^{L}\left[\sigma_{i}^{x}\sigma_{i+1}^{x}+h\sigma_{i}^{z}-\kappa\sigma_{i}^{x}\sigma_{i+2}^{x}\right]
\]
and perform numerically simulated quenches from $(h_{0},\kappa_{0})$
to $(h_{1},\kappa_{1})$. At $\kappa=0$ the model can be mapped to
quasi-free fermions while for any non zero $\kappa$ the Hamiltonian
is not integrable. As shown in fig.~\ref{fig:var_kappa}, the variance
of $m\left(t\right)=\langle\sigma_{i}^{z}\left(t\right)\rangle$ is
discontinuous at the integrable point $\kappa=0$. As expected $\Delta m^{2}\left(\kappa=0\right)$
is larger than $\Delta m^{2}\left(\kappa\to0\right)$. We would like
to stress here that the appearance of a discontinuity in an infinity
time average, is perfectly legitimate even for finite size systems.
The origin of the discontinuity of the variance stems from a massive
violation of the non-resonant condition at $\kappa=0$. Note that
the average $\overline{m}$ (not shown) is smooth at $\kappa=0$,
indicating that the degeneracies of the energies are constant around
$\kappa=0$. We also performed numerical simulations keeping the observation
time window $T$ finite. The effect of a finite $T$ is to make the
variance smooth at $\kappa=0$, approaching the discontinuous $T=\infty$
value with corrections of order $T^{-1}$.

As noted previously, it is easy to prove that at the integrable point
the variance scale as $\Delta m^{2}=O\left(L^{-1}\right)$. On the
contrary, checking the exponentially small scaling expected at non-integrable
points is very difficult numerically as the computation of the variance
requires full exact diagonalization which limits the analysis to very
short sizes. 

Finally, a compelling question is whether similar results generalize
to more complex integrable models such as those integrable by Bethe
Ansatz. To investigate this scenario we performed preliminary numerical
simulations with the following Hamiltonian
\begin{align}
H & =\sum_{i=1}^{L}\left[\sigma_{i}^{x}\sigma_{i+1}^{x}+\sigma_{i}^{y}\sigma_{i+1}^{y}+\Delta\sigma_{i}^{z}\sigma_{i+1}^{z}+\right.\nonumber \\
 & \left.\alpha\left(\sigma_{i}^{x}\sigma_{i+2}^{x}+\sigma_{i}^{y}\sigma_{i+2}^{y}+\Delta\sigma_{i}^{z}\sigma_{i+2}^{z}\right)\right],\label{eq:H_alpha}
\end{align}
with periodic boundary condition. At $\alpha=0$ the Hamiltonian is
integrable by Bethe Ansatz. In our simulations we quenched from $\left(\Delta_{0},\alpha_{0}\right)$
to $\left(\Delta_{1},\alpha_{1}\right)$, and looked at the statistics
of $\mathcal{A}\left(t\right)=\langle\sigma_{i}^{+}\sigma_{i+1}^{-}\left(t\right)\rangle+\hc$
as this is a quadratic observable in the fermionic setting. Our numerics
shows that the fluctuations are smooth as $\alpha_{1}$ crosses the
integrable point, which indicates that the non-resonant condition
is satisfied (when restricted to the relevant subspace) also at the
integrable point. This indeed has to be expected from the form of
the many-body energies arising from the Bethe Ansatz %
\footnote{V. Gritsev, private communication.%
}. However it is still possible that a lack of rational independence
of the many-body energies, for Bethe Ansatz integrable models, will
show up in a non-analytic behavior of higher cumulants of $\mathcal{A}\left(t\right)$
as a function of the non-integrability parameter $\alpha$. 

\begin{figure}
\begin{centering}
\includegraphics[width=7cm]{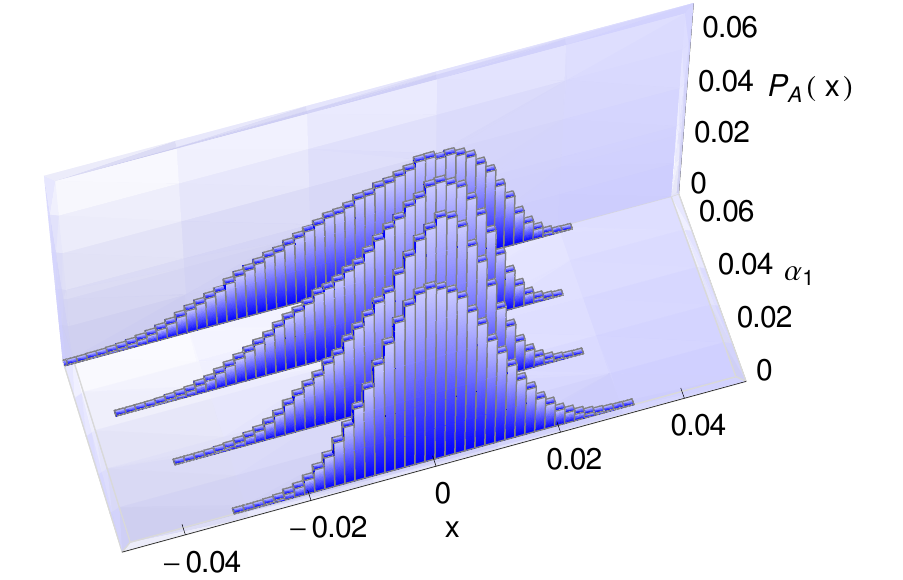}
\par\end{centering}

\caption{Full time statistics $P_{A}\left(x\right)$ for the observable $\mathcal{A}\left(t\right)=\langle\sigma_{i}^{+}\sigma_{i+1}^{-}\left(t\right)\rangle+\hc$
for the Hamiltonian Eq.~(\ref{eq:H_alpha}). The quench parameters
are $\left(\Delta_{0}=2,\,\alpha_{0}=1.7\right)\to\left(\Delta_{1}=1,\alpha_{1}=0,0.02,0.04,0.06\right)$
for a size $L=10$. At $\alpha_{1}=0$ the Hamiltonian is integrable
by Bethe-Ansatz. Data are obtained with full diagonalization of Hamiltonian
Eq.~(\ref{eq:H_alpha}). For the time statistics, the observable
$\mathcal{A}\left(t\right)$ has been computed at 200,000 random times
distributed uniformly in $\left[0,T\right]$ with $T=100,000$.\label{fig:Full-time-statistics}}
\end{figure}

In figure \ref{fig:Full-time-statistics} we plot the distribution
function $P_{A}$ for $\mathcal{A}\left(t\right)$ for various quench
experiment. At the integrable point the distribution looks qualitatively
different from the distribution obtained at $\alpha_{1}\neq0$. However
further investigations are needed to ascertain whether the dependence
of $P_{A}$ on $\alpha_{1}$ is analytic at the integrable point $\alpha_{1}=0$.
A lack of analyticity would be a signature of a lack of rational independence
of the many-body energies. 

\section{Conclusions}

In this paper we addressed the question of equilibration in quasi-free
Fermi systems. The initial state is a general state while the evolution
Hamiltonian as well as the observable are quadratic in the Fermi operators.
While for general non-free systems the variance $\Delta\mathcal{A}^{2}$
is typically exponentially small in the volume, we find that in the
quasi-free setting (for extensive observables) both the mean and the
variance are proportional to the volume. This hints at the possibility
of an underlying central limit theorem, a circumstance that we termed
\emph{Gaussian equilibration}. In this case the properly rescaled
observable becomes Gaussian in the large $L$ limit, and in general
the relative error satisfies $\Delta\mathcal{A}/\overline{\mathcal{A}}\sim1/\sqrt{L}$.
We proved Gaussian equilibration for the magnetization in the quantum
XY model assuming rational independence of the one body energies and
gave evidence for a class of observables/initial states evolving with
a tight-binding model. In all cases Gaussian equilibration was confirmed
by numerical simulations. The key insight is a mapping of the equilibration
dynamics to a generalized classical XY model at infinite temperature.
As a by product we obtain a quantum setting (initial state, Hamiltonian,
and observable $A_{XY}$), such that the equilibration dynamics of
$A_{XY}$ gives the solution of a classical XY model in $D$-dimension,
and vice-versa. 

A consequence of the above scenario --confirmed by numerical simulations--
is that the variance $\Delta\mathcal{A}^{2}$ turns out to be discontinuous
at the quasi-free point of an otherwise non-integrable Hamiltonian.
This shows that the enhancement of the temporal fluctuations of $\Delta\mathcal{A}^{2}$
(for a quadratic observable $A$) may provide a universal and \emph{experimentally
testable} signature of integrability in the context of out of equilibrium
dynamics, at least when integrable systems are identified with quadratic
models. 

Ultimately the origin of the exponential decrease of the signal to
noise ratio illustrated in this paper lies in the exponential reduction
of the effective phase space entailed by the quasi-free nature of
our setup i.e., the many body space gets effectively replaced by the
single particle one. A natural question for future research is whether
a similar mechanism is at place for a more general class of integrable
systems such as Bethe-Anstanz integrable models. Preliminary numerical
results indicate that, in this case the variance is smooth and a possible
non-analytic behavior at the integrable point must be sought in higher
order cumulants. 

LCV wishes to thank KITP for the kind hospitality. This research was
partially supported by the ARO MURI under grant W911NF-11-1-0268 and
by the National Science Foundation under Grant No.~NSF PHY11-25915.
PZ also acknowledges partial support by NSF grants No.~PHY-969969
and No.~PHY-803304.

\bibliography{ge}

\end{document}